\documentclass[10pt,a4paper,final]{article}
\usepackage{graphicx}
\usepackage[breaklinks=true,colorlinks=true,linkcolor=blue,urlcolor=blue,citecolor=blue]{hyperref}
\usepackage{cite}
\usepackage{amsmath}
\usepackage{braket} 
\usepackage{tikz}
\usetikzlibrary{quantikz}
\usepackage{bbold}
\usepackage{comment}
\usepackage{xcolor}
\usepackage{geometry}
\newgeometry{vmargin={25mm}, hmargin={25mm,25mm}}   

\newcommand{\eeq}{\end{equation}}
\newcommand{\beq}{\begin{equation}}
\newcommand{\no}{\noindent}

\newcommand{\cnbls}{$CNOT$ blocks\hspace{0.5em}}
\newcommand{\Tr}{\text{Tr}}

\begin{document}

\title{Escaping barren plateaus in approximate quantum compiling}
\date{}

\maketitle

\begin{center}
{\large Niall F. Robertson$^{1}$, Albert Akhriev$^{1}$, Jiri Vala$^{2,3}$ and Sergiy Zhuk$^{1}$}
\end{center}

\vspace{0.5cm}

{\no \sl\small $^1$ IBM Quantum, IBM Research Europe - Dublin, IBM Technology Campus, Dublin 15, Ireland\\}
{\sl\small $^2$ Maynooth University, Maynooth, Ireland\\}
{\sl\small $^3$ Tyndall National Institute, Cork, Ireland}

\begin{abstract}
Quantum compilation provides a method to translate quantum algorithms at a high level of abstraction into their implementations as quantum circuits on real hardware. One approach to quantum compiling is to design a parameterised circuit and to use techniques from optimisation to find the parameters that minimise the distance between the parameterised circuit and the target circuit of interest. While promising, such an approach typically runs into the obstacle of barren plateaus - i.e. large regions of parameter space in which the gradient vanishes. A number of recent works focusing on so-called quantum assisted quantum compiling have developed new techniques to induce gradients in some particular cases. Here we develop and implement a set of related techniques such that they can be applied to classically assisted quantum compiling. We consider both approximate state preparation and approximate circuit preparation and show that, in both cases, we can significantly improve convergence with the approach developed in this work.
\end{abstract}

\section{Introduction}
\begin{figure}[h]
\centering
\begin{tikzpicture}[scale = 1.5]
\draw[gray, thick] (-0.3,1) -- (0.3,1);
\draw[gray, thick] (0,1) -- (0,0);
\filldraw[black] (0,1) circle (2pt);
\draw[black] (0,0.12) circle (4pt);
\draw[gray, thick] (-0.3,0.13) -- (0.3,0.13);

\draw[gray, thick] (4,1) -- (4,0);
\draw[gray, thick] (3.7,0) -- (4.3,0);
\filldraw[black] (4,0) circle (2pt);
\draw[black] (4,0.88) circle (4pt);
\draw[gray, thick] (3.7,0.87) -- (4.3,0.87);
\end{tikzpicture}
\caption{Controlled-NOT ($CNOT$) operations with the role of a control qubit played by the first and second qubit respectively. $CNOT$ transforms each standard computational basis element in a quantum superposition by adding a value of the control qubit to that of the target qubit modulo 2: $\ket{a}_c \otimes \ket{b}_t \rightarrow \ket{a}_c \otimes \ket{b \oplus a\mod2}_t$ where $a,b \in \{0, 1\}$.}
\label{CNOTs}
\end{figure}
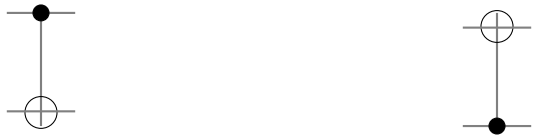

Quantum compiling is a set of tools to translate a quantum circuit into a set of gates that can be efficiently implemented on quantum hardware. To do so, one must take into account hardware constraints such as connectivity and maximum circuit depth. Limited hardware connectivity between qubits requires additional operations such as $SWAP$ gates to be included into the compilation process \cite{nannicini2021optimal}.  Similarly, the physical implementation of entangling two-qubit operations such as a controlled-NOT ($CNOT$) operation, illustrated in Fig. \ref{CNOTs}, may require that only one of the two qubits can play the role of the control qubit. Despite recent progress on error mitigation techniques \cite{temme2017error, kim2021scalable, giurgica2020digital}, noise on present day devices significantly limits the depth of the circuit that can be implemented - the benefit of optimally decomposing a quantum circuit into the hardware's native gate alphabet and connectivity constraints is hence self-evident. We can classify quantum compiling techniques into two broad categories: classically assisted and quantum assisted. In the former case, the idea is to use classical methods to compile sub-circuits that form part of a larger quantum algorithm. The main advantage of such an approach is that one can avoid the issues arising from noise on a quantum device, the disadvantage being that, due to the exponential scaling of the Hilbert space, one is limited to sub circuits that act on a fixed number of qubits. The quantum assisted approach however is scalable and can hence be applied to the entire quantum circuit of interest. However, one encounters quantum noise which, even in cases where the compilation scheme is somewhat noise resilient \cite{sharma2020noise}, can cause significant convergence issues \cite{wang2021noise}. Furthermore, the circuits that have been proposed to implement a quantum assisted quantum compilation procedure require long range qubit connectivity on the quantum device of interest \cite{khatri2019quantum}.\\

Here we focus on the classically assisted approach to quantum compilation. Classically assisted variational quantum algorithms have been recently studied in \cite{rudolph2022synergy, rudolph2022decomposition}. In these works, Tensor Networks are used to generate a shallow circuit with weak entanglement. The circuit is subsequently deepened on a quantum device, and the entanglement is thus increased beyond what classical techniques can store. These recent promising results provide significant motivation to improve the state of the art of classical optimisation techniques applied to quantum circuits which are currently limited by convergence issues such as barren plateaus, as will be discussed below.\\

We consider the problem of \textit{approximate quantum compiling} as studied in \cite{madden2022best}. This approach involves the design of a parametric quantum circuit with fixed depth - the parameters are then adjusted to bring it as ``close" as possible to the target, where ``close" is defined via some carefully chosen metric (see section \ref{sec:opt}). In \cite{madden2022best}, it was shown that such an approach could be used to derive a 14-$CNOT$ decomposition of a 4-qubit Toffoli gate from scratch. It was furthermore shown that the Quantum Shannon Decomposition, introduced by Shende, Bullock and Markov \cite{Shende_06}, can be compressed by a factor of two without practical loss of fidelity. In this work, our main contribution is a new classical algorithm for quantum compilation which significantly improves convergence - see below and main text for details.\\

In addition to the sub-circuit problem discussed above, we consider state preparation as an example where classical techniques can play a significant role. For a large class of quantum algorithms, the entanglement of the wavefunction is weakly entangled at the beginning of the circuit and grows with the depth. There exist widely studied classical techniques (i.e. Tensor Networks) to store the wavefunction of a large number of qubits when the system is only weakly entangled. Thus, this opens the possibility of using classical Tensor Network techniques for state preparation to optimally compile the early parts of a quantum circuit, in a similar way to \cite{rudolph2022synergy, rudolph2022decomposition}. One of the biggest obstacles however to any quantum compilation technique, whether it is quantum assisted or not, is the existence of barren plateaus in the parameter landscape, i.e. large regions where the gradient of the cost function vanishes exponentially in the number of qubits, thus rendering it extremely difficult to train. A proposed solution for the quantum assisted quantum compiling approach is to use shallow circuit Ansatz' and so-caled ``local cost functions". In short, a cost function calculated via a quantum circuit is local if it involves the measurement of only a small number of qubits - a global cost function involves the measurement of all qubits. Here we design a new classical algorithm inspired by the distinction between local and global cost functions made in \cite{khatri2019quantum, cerezo2021cost} for quantum compilation. In particular, we derive an expression inspired by the quantum approach that can be easily implemented on a classical computer and that allows us to escape the barren plateaus. This new cost function introduces a number of coefficients that we must update throughout the optimsation procedure. Our algorithm implements a scheme to do so effectively, such that short depth parametric quantum circuits can be found that closely approximate the target circuit of interest.\\

Our main contributions in this work are as follows: we derive an explicit expression for the local cost functions considered in \cite{khatri2019quantum, cerezo2021cost} in equation (\ref{localcost}), such that they obtain a clear and intuitive operational meaning. We use this to develop a new cost function - see equation (\ref{cost_func_alpha}) - whose gradient can be efficiently calculated classically. We consider a particular case where our cost function can be studied analytically and we show how the variance of its gradient behaves - see equation (\ref{var_clk}) - which allows us to understand how and why our classical compilation scheme allows for the escape from barren plateaus. We develop an algorithm to update the coefficients appearing in our new cost function - see section \ref{sec:algorithm}, appendix \ref{weight_scheme} and appendix \ref{surrogate_mod}. Finally, we demonstrate numerically in Figures \ref{fig:lhs-vs-no-lhs} and \ref{8q} that the combination of our cost function and our algorithm to update the coefficients significantly improves the convergence of certain problems in quantum compilation.\\

The outline of this article is as follows: in section \ref{sec:setup}, we review approximate quantum compiling and we show how it can be formulated as an optimisation problem. In section \ref{sec:barren}, we discuss the convergence issues that are commonly encountered in quantum compiling problems - we discuss in particular the appearance of barren plateaus in the parameter landscape. We introduce our classical version of the local cost function and we show how, for a particular example where the cost function can be studied analytically, the variance of the gradient of our cost function behaves as a function of the number of qubits. In section \ref{sec:algorithm}, we describe our scheme to update the coefficients that appear in our classical local cost function over the course of the optimisation procedure. In section \ref{sec:results}, we present some numerical results on simulations for up to 24 qubits showing that our classical local cost function significantly improves convergence as compared to the global one. We conclude in section \ref{sec:discussion}.

\section{Setup}\label{sec:setup}

Quantum compilation is an active research area with a wide spectrum of research directions. One such direction is to find the the best representation - in terms of some distance or structural properties - of a generic unitary in a certain canonical basis (e.g.~\cite{kliuchnikov2012fast, ross2014, sarnak2015letter, selinger2013nqubit, vala2004}). One can also consider an approach based on exact qubit routing, i.e. mapping logical qubits to physical qubits so that the resulting unitary coincides with the original one (up to a permutation) while allowing for compatibility with the hardware topology (e.g.~\cite{maslov2008a, amy2013, nam2018, maslov2008b, bhattacharjee2017, oddi2018, booth2018, zulehner2019a, bhattacharjee2019, zulehner2019b, cowtan2019, tket2020, itoko2020, tan2020optimality}). As discussed in the introduction, the approach to quantum compilation that we focus on here is to design a parametric quantum circuit with fixed depth and to adjust the parameters of this circuit to bring it as ``close" as possible to the target, where ``close" is defined via some carefully chosen metric. While one can in principle use universal parametric circuits which provide a theoretical guarantee that there exist parameters such that any target circuit can be exactly compiled, such circuit templates result in impractically deep circuits for present day quantum devices. Approximate quantum compiling can be seen as a compilation approach that trades off the circuit accuracy for its depth. It was shown in \cite{knill1995approximation} that generic unitary operations over $n$ qubits are computationally hard even to approximate. More recently it has been shown that quantum circuit compilation is an NP-complete problem \cite{botea2018complexity}. However, the fact that there is a subset of unitary operations that can be realized in quantum circuits of lower complexity, such as those in the Shor algorithm \cite{shor1999polynomial} for example, suggests that quantum compilation may offer both an interesting perspective on complexity of quantum circuits and perhaps a path to new quantum algorithms that would provide quantum advantage. 
\subsection{Parametric quantum circuits and ansatz}\label{sec:param_circ}

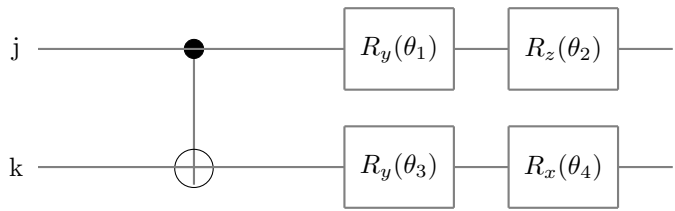
\begin{figure}[h]
\centering
\begin{tikzpicture}[scale = 1.8]
\draw[gray, thick] (0,1) -- (0,0);
\filldraw[black] (0,1) circle (2pt);
\draw[black] (0,0.12) circle (4pt);
\draw[gray, thick] (-0.14,0.13) -- (0.14,0.13);

\draw[gray, thick] (-1.14,1) -- (1.1,1);
\draw[gray, thick] (-1.14,0.13) -- (1.1,0.13);

\draw[gray, thick] (1.1,0.7) -- (1.1,1.3);
\draw[gray, thick] (1.9,0.7) -- (1.9,1.3);
\draw[gray, thick] (1.1,1.3) -- (1.9,1.3);
\draw[gray, thick] (1.1,0.7) -- (1.9,0.7);

\draw[gray, thick] (1.1,-0.17) -- (1.1, 0.43);
\draw[gray, thick] (1.9,-0.17) -- (1.9, 0.43);
\draw[gray, thick] (1.1, 0.43) -- (1.9, 0.43);
\draw[gray, thick] (1.1,-0.17) -- (1.9,-0.17);

\draw[gray, thick] (1.9,1) -- (2.3,1);
\draw[gray, thick] (1.9,0.13) -- (2.3,0.13);

\draw[gray, thick] (2.3,-0.17) -- (2.3, 0.43);
\draw[gray, thick] (3.1,-0.17) -- (3.1, 0.43);
\draw[gray, thick] (2.3, 0.43) -- (3.1, 0.43);
\draw[gray, thick] (2.3,-0.17) -- (3.1,-0.17);

\draw[gray, thick] (2.3,0.7) -- (2.3,1.3);
\draw[gray, thick] (3.1,0.7) -- (3.1,1.3);
\draw[gray, thick] (2.3,1.3) -- (3.1,1.3);
\draw[gray, thick] (2.3,0.7) -- (3.1,0.7);

\draw[gray, thick] (3.1,1) -- (3.5,1);
\draw[gray, thick] (3.1,0.13) -- (3.5,0.13);

\node at (-1.3,1) {j};
\node at (-1.3,0.13) {k};

\node at (1.5,1) {$R_y(\theta_1)$};
\node at (2.7,1) {$R_z(\theta_2)$};
\node at (1.5,0.13) {$R_y(\theta_3)$};
\node at (2.7,0.13) {$R_x(\theta_4)$};

\end{tikzpicture}
\caption{$CNOT$ block forms the basic building block of our circuit ansatz.}\label{CNOTblock}
\end{figure}

As discussed in \cite{madden2022best}, a natural circuit Ansatz is one based on so-called $CNOT$ blocks. A $CNOT$ block is a $CNOT$ gate followed by single qubit rotations (see Figure \ref{CNOTblock}). Single-qubit rotation gates are defined as complex exponential functions of the Pauli matrices:
\beq
\begin{aligned}
&R_{x}(\theta)=\exp (-i \sigma_x \theta / 2)=\left[\begin{array}{cc}
\cos (\theta / 2) & -i \sin (\theta / 2) \\
-i \sin (\theta / 2) & \cos (\theta / 2)
\end{array}\right], \\
&R_{y}(\theta)=\exp (-i \sigma_y \theta / 2)=\left[\begin{array}{cc}
\cos (\theta / 2) & -\sin (\theta / 2) \\
\sin (\theta / 2) & \cos (\theta / 2)
\end{array}\right], \\
&R_{z}(\theta)=\exp (-i \sigma_z \theta / 2)=\left[\begin{array}{cc}
\exp (-i \theta / 2) & 0 \\
0 & \exp (i \theta / 2)
\end{array}\right].
\end{aligned}
\eeq
\\
\begin{figure}
\centering
  \begin{quantikz}
      &  \gate{R_z(\theta_1)} & \gate{R_y(\theta_2)} & \gate{R_z(\theta_3)} & \gate[wires=2]{}& \gate[wires=2]{}&\gate[wires=2]{} & \qw & \qw & \qw & \gate[wires=2]{} & \gate[wires=2]{} & \gate[wires=2]{} & \qw & \qw & \qw & \qw &  \\
       & \gate{R_z(\theta_4)} & \gate{R_y(\theta_5)} & \gate{R_z(\theta_6)}& \qw & \qw  & \qw  & \gate[wires=2]{}& \gate[wires=2]{}& \gate[wires=2]{}& \qw & \qw & \qw & \gate[wires=2]{} & \gate[wires=2]{} & \gate[wires=2]{} & \qw &   \\
       &  \gate{R_z(\theta_7)} & \gate{R_y(\theta_8)} & \gate{R_z(\theta_9)} & \gate[wires=2]{}& \gate[wires=2]{}&\gate[wires=2]{}& \qw & \qw & \qw & \gate[wires=2]{} & \gate[wires=2]{} & \gate[wires=2]{} & \qw & \qw & \qw & \qw & \\
       & \gate{R_z(\theta_{10})} & \gate{R_y(\theta_{11})} & \gate{R_z(\theta_{12})} & \qw & \qw & \qw & \qw & \qw & \qw & \qw & \qw & \qw & \qw & \qw & \qw & \qw &\\
  \end{quantikz}
  \caption{Parameterised circuit with $n=4$ qubits, $l=2$ layers and $b=3$ CNOT blocks in each layer.}\label{paramcircuit}
\end{figure}
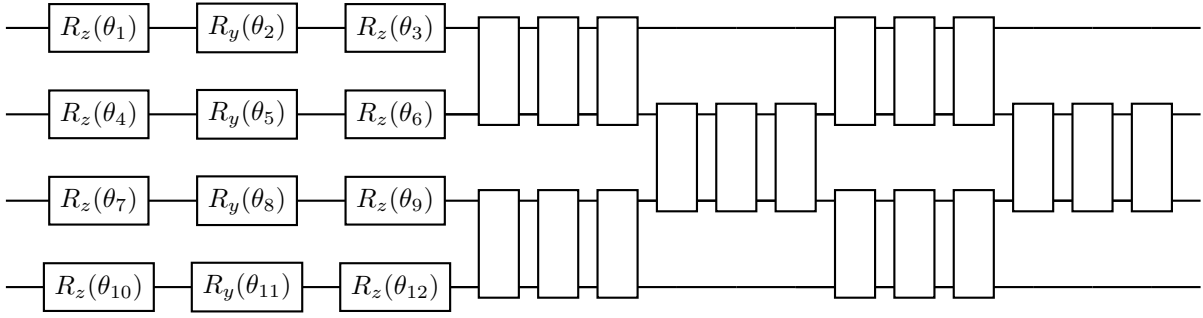
A block with a $CNOT$ gate acting on a ``control" qubit $j$ and ``target" qubit $k$ is written as $\mathrm{CU}_{jk}(\theta_1, \theta_2, \theta_3, \theta_4)$. For a given hardware connectivity, one can then write down a full parameterised circuit as: 
\beq
\begin{aligned}
V_{\mathrm{ct}}(\boldsymbol{\theta})=& \mathrm{CU}_{\mathrm{ct}(L)}\left(\theta_{3 n+4 L-3}, \ldots, \theta_{3 n+4 L}\right) \cdots \mathrm{CU}_{\mathrm{ct}(1)}\left(\theta_{3 n+1}, \ldots, \theta_{3 n+4}\right) \\
& {\left[R_{z}\left(\theta_{1}\right) R_{y}\left(\theta_{2}\right) R_{z}\left(\theta_{3}\right)\right] \otimes \cdots \otimes\left[R_{z}\left(\theta_{3 n-2}\right) R_{y}\left(\theta_{3 n-1}\right) R_{z}\left(\theta_{3 n}\right)\right] }
\end{aligned}
\eeq
The position of the $CNOT$ blocks in the parameterised circuit can be customised to suit the particular target circuit that one is interested in. We will use a particular structure of the ansatz which is illustrated in Figure \ref{paramcircuit}. In this circuit Ansatz, we have 3 rotation angles for each qubit at the beginning, we then have $l$ layers in which we repeat each $CNOT$ block $b$ times. In Figure \ref{paramcircuit}, we have $n=4$ qubits, $l=2$ and $b=3$. We note that for $b=3$, each block of $CNOT$s acting on $2$ qubits constitutes a universal two-qubit quantum circuit which represents an arbitrary unitary rotation of a two-qubit quantum state.\\

\noindent For an $n$-qubit circuit, we typically take the initial state to be $\ket{0}^{\otimes n}$ which, in the illustrated example, is first transformed by a sequence of three single-qubit rotations $R_z(\theta_1)$, $R_y(\theta_2)$, $R_z(\theta_3)$. Strictly speaking, only two inequivalent single-qubit rotations are required if the initial state is the fiducial initial state $\ket{\boldsymbol{0}}$ but we use three for implementation in order to consider more general situations \cite{madden2022best}.

\subsection{Optimization problem}\label{sec:opt}

The compilation now becomes an optimisation problem whereby one aims to tune the parameters of the rotation gates to minimise the distance between the parametric circuit and the target. Given a target unitary matrix $U \in \mbox{U}(2^n)$ and a set of constraints, such as connectivity and length , the approximate quantum compiling problem is to find the unitary matrix $V(\theta) \in \mbox{U}(2^n)$, that can be implemented with parametrized quantum circuit and that is closest distance to the target $U$. One can define this distance via a number of metrics such as the Frobenius norm
\beq\label{frobenius}
\min _{V \in U\left(2^{n}\right)} f(V):=\frac{1}{2}\|V(\theta)-U\|_{\mathrm{F}}^{2} = 1 - \frac{1}{2^{n}} \mbox{Re} [\Tr(V^\dagger(\theta) U) ]
\eeq
Similarly, one can also use the Hilbert-Schmidt test:
\beq\label{hstest}
C_{HS} = 1 - \frac{1}{d^2}|\Tr(V^{\dagger}(\theta)U)|^2
\eeq
where $d$ is the dimension of the Hilbert space. The Hilbert-Schmidt test (\ref{hstest}) is closely related to the Frobenius norm in equation (\ref{frobenius}) and used as the cost function in \cite{khatri2019quantum}. We have:
\beq\label{frob_real}
C_{FN}\equiv \frac{1}{2}\|V(\theta)-U\|_{\mathrm{F}}^{2} = d - \text{Re} \left[\Tr(V^{\dagger}(\theta)U)\right]
\eeq
From (\ref{hstest}) and (\ref{frob_real}), we get:
\beq
1-C_{HS} = \left(1-\frac{C_{FN}}{d}\right)^2 + \frac{1}{d^2}\left(\text{Im} (\Tr(V^{\dagger}U))\right)^2
\eeq
and hence:
\beq
1-C_{HS} \geq \left(1-\frac{C_{FN}}{d}\right)^2
\eeq
We can thus conclude that minimising $C_{FN}$ minimises $C_{HS}$ automatically. We will show in this work that one can improve the convergence issues caused by barren plateaus in approximate quantum compiling by minimising a cost function that is related to (\ref{hstest}) above but also includes some additional perturbation terms that ``localise" the original cost function - a concept that will be precisely defined below.

\section{Barren Plateaus}\label{sec:barren}

\no The Frobenius norm in equation (\ref{frobenius}) provides a natural cost function for Approximate Quantum Compiling. In particular, one can indeed see that this quantity is $0$ if and only if $V(\theta) = U$. However, this cost function is problematic in practice and is only feasible for a small number of qubits $n$ due to the appearance of barren plateaus, i.e. large regions of parameter space where the gradient of the cost function vanishes exponentially, thus rendering it impossible to train the cost function. Not only is this observed in practice, but a number of recent works \cite{cerezo2021cost ,wang2021noise, holmes2022connecting} have even provided strict proofs that guarantee the existence of these barren plateaus in a variety of circumstances, where the definition of a barren plateau for these proofs is that the variance of the gradient vanishes exponentially with the number of qubits $n$. We remark that these works were focused on quantum assisted quantum compiling in which case an exponentially vanishing gradient requires an exponential number of measurements to estimate the gradient. In the classical implementation of quantum compiling that we consider here, we calculate the gradient directly without taking measurements and thus in principle an exponentially vanishing gradient is not as significant of an issue. However, in practice we observe that even in the classical case, the appearance of these barren plateaus cause the training times to increase beyond practical limits. Furthermore, we observe that the barren plateaus appear even in systems with small numbers of qubits $n$. We hence require new tools to overcome these issues. \\

\begin{figure}
\centering
  \begin{quantikz}
  \lstick[wires = 4]{A} & 
     & \lstick{$\ket{0}$} & \gate{H} & \ctrl{4} &  \qw & \qw & \gate[wires=4, nwires={3}]{U} & \qw & \qw & \qw & \ctrl{4} & \gate{H} & \meter{}\\
    & & \lstick{$\ket{0}$} & \gate{H} & \qw & \ctrl{4} &  \qw & \qw & \qw  & \qw & \ctrl{4} & \qw & \gate{H} & \meter{} \\
    & & \lstick{$\vdots$} &  &  &  &  & & \\
    & & \lstick{$\ket{0}$} & \gate{H} & \qw & \qw & \ctrl{4} & \qw & \qw &\ctrl{4} & \qw  & \qw & \gate{H} & \meter{}\\
     \lstick[wires=4]{B} & & \lstick{$\ket{0}$} & \qw & \targ{} & \qw &  \qw & \gate[wires=4, nwires={3}]{V^*}  & \qw & \qw & \qw  & \targ{} & \qw & \meter{} \\
    & & \lstick{$\ket{0}$} & \qw & \qw & \targ{} & \qw & \qw & \qw & \qw & \targ{} & \qw & \qw & \meter{} \\
    & & \lstick{$\vdots$} \\
   &  & \lstick{$\ket{0}$} & \qw & \qw & \qw & \targ{} & \qw & \qw & \targ{} & \qw  & \qw & \qw & \meter{}\\
  \end{quantikz}
  \caption{The quantum circuit discussed in \cite{khatri2019quantum} that is used to evaluate the expression in (\ref{cost_state_prep}).}\label{global}
\end{figure}
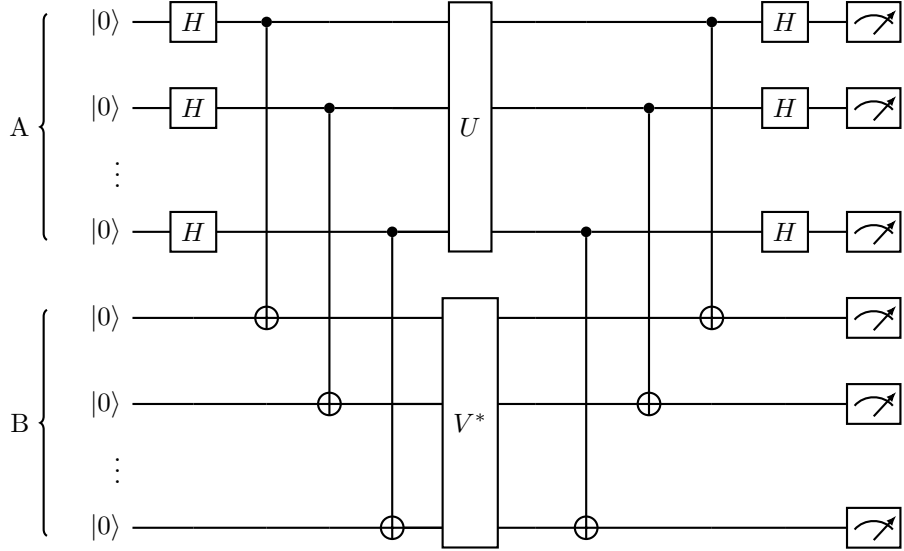

In \cite{khatri2019quantum, cerezo2021cost}, it was shown that one can escape the barren plateau with so-called \textit{local} cost functions as opposed to \textit{global} cost functions - to be defined below - as long as the circuit ansatz $V(\theta)$ is sufficiently shallow. In particular, as shown in \cite{cerezo2021cost}, if the depth of $V(\theta)$ is of order $\mathcal{O}(\log n)$ then the variance of the gradient of a local cost function vanishes only polynomially instead of exponentially. Broadly speaking, a global cost function is one that is evaluated by measuring all qubits at once, whereas a $k-$local cost function is evaluated by measuring only $k$-qubits at once. More specifically, in the context of state preparation circuits as studied in \cite{cerezo2021cost}, a global cost function is defined as one of the form:
\beq
C_{\mathrm{G}}=\operatorname{Tr}\left[O_{\mathrm{G}} V(\boldsymbol{\theta})\left|\psi_{0}\right\rangle\left\langle\psi_{0}\right| V(\boldsymbol{\theta})^{\dagger}\right]
\eeq
where $O_{\mathrm{G}}$ is an operator that acts non-trivially on all qubits and $\ket{\psi_0}$ is the target state. An example would be:
\beq
\quad O_{\mathrm{G}}=\mathbb{1}-\ket{\boldsymbol{0}}\bra{\boldsymbol{0}},
\eeq
A local cost function, on the other hand, is defined as:
\beq
C_{\mathrm{L}}=\operatorname{Tr}\left[O_{\mathrm{L}} V(\boldsymbol{\theta})|\psi_{0}\rangle\langle\psi_{0}| V(\boldsymbol{\theta})^{\dagger}\right]
\eeq
where $O_{\mathrm{L}}$ is a sum of operators which act non-trivially on a local subset of qubits. An example would be:
\beq
\quad O_{\mathrm{L}}=\mathbb{1}-\frac{1}{n} \sum_{j=1}^{n}\ket{0}\bra{0}_{j} \otimes \mathbb{1}_{\bar{j}}.,
\eeq
where $\bar{j}$ is the complement of the $j$-th qubit. In \cite{khatri2019quantum} the evaluation of the cost function in (\ref{hstest}) on a quantum computer was discussed. In particular, it was shown that for a target circuit $U$ and approximate circuit $V(\theta)$ acting on $n$ qubits, one can calculate the quantity $\frac{1}{d^2}|\Tr(V^{\dagger}(\theta)U)|^2$ in equation (\ref{hstest}) by running the circuit in Figure \ref{global} - which acts on $2n$ qubits - and measuring the probability of observing all zeros at the end of the circuit. Since this circuit involves the measurement of all qubits the cost function is classified as a global cost function and hence suffers from the barren plateau problems discussed above. \\

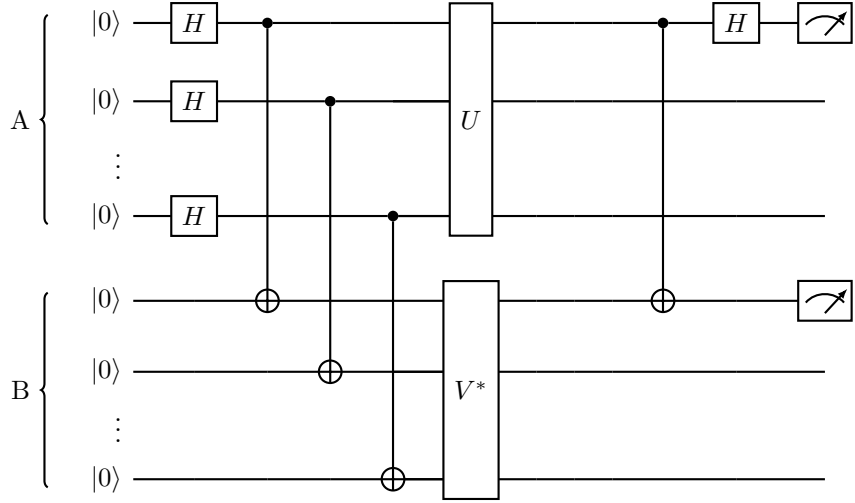
\begin{figure}
\centering
  \begin{quantikz}
  \lstick[wires = 4]{A} & & \lstick{$\ket{0}$} & \gate{H} & \ctrl{4} &  \qw & \qw & \gate[wires=4, nwires={3}]{U} & \qw & \qw & \qw & \ctrl{4} & \gate{H} & \meter{}\\
    & & \lstick{$\ket{0}$} & \gate{H} & \qw & \ctrl{4} &  \qw & \qw & \qw  & \qw & \qw & \qw & \qw & \qw\\
    & & \lstick{$\vdots$} &  &  &  &  & & \\
    & & \lstick{$\ket{0}$} & \gate{H} & \qw & \qw & \ctrl{4} & \qw & \qw &\qw & \qw  & \qw & \qw & \qw \\
     \lstick[wires=4]{B} & & \lstick{$\ket{0}$} & \qw & \targ{} & \qw &  \qw & \gate[wires=4, nwires={3}]{V^*}  & \qw & \qw & \qw  & \targ{} & \qw & \meter{} \\
    & & \lstick{$\ket{0}$} & \qw & \qw & \targ{} & \qw & \qw & \qw & \qw & \qw & \qw & \qw & \qw \\
    & & \lstick{$\vdots$} &  &  &  &  & & & \\
   &  & \lstick{$\ket{0}$} & \qw & \qw & \qw & \targ{} & \qw & \qw & \qw & \qw  & \qw & \qw & \qw\\
  \end{quantikz}
  \caption{The quantum circuit discussed in \cite{khatri2019quantum} that is used to evaluate the local version of the Hilbert-Schmidt test.}\label{local}
\end{figure}

In \cite{khatri2019quantum}, the authors describe a natural way to create a local version of the Hilbert-Schmidt cost function. One starts with the circuit in Figure \ref{global} and replaces the measurements on all qubits with a measurement of qubits $j$ and $j+n$, as shown in Figure \ref{local}, and measures the probability $p_0^j$ of observing zeros on both qubits. The local cost function is then obtained by taking the average of all the $p_0^j$'s. In what follows, we will explicitly calculate an expression for this quantity such that it has an intuitive operational meaning.\\

As discussed in the introduction, another case for which a classical implementation of approximate quantum compiling is useful is when the target circuit appears at the beginning of the circuit. In this case, the problem essentially becomes one of state preparation where the target state is given by $\ket{\psi_0} = U\ket{0}$. One can use classical techniques to store the wavefunction $\ket{\psi_0}$ of the system even for large numbers of qubits if the entanglement is low enough \cite{rudolph2022synergy, rudolph2022decomposition}. For some quantum circuits, such as those that simulate the time evolution of quantum spin chains after a quantum quench \cite{smith2019simulating}, the initial state is weakly entangled and grows linearly with the depth of the circuit. For such circuits, our classical implementation of approximate quantum compiling can be applied to a sub-circuit at the beginning of the algorithm when the entanglement is low. The Hilbert-Schmidt test in (\ref{hstest}) applied to state preparation becomes:
\beq\label{cost_state_prep}
C_{hs}^{\text{state}} = 1 -  |\bra{0}V^{\dagger}(\theta)\ket{\psi_0}|^2 .
\eeq
The cost function in (\ref{cost_state_prep}) is now equal to the probability of observing of all zeros after preparing the state $\ket{\psi_0}$ on the quantum computer followed by the application of $V^{\dagger}$. The local version of (\ref{cost_state_prep}) becomes:
\beq\label{locprep}
C_{\text{local}} = 1-\frac{1}{n}\sum\limits_{j=1}^{n}p_0^j
\eeq
where $p_0^j$ is the probability of observing zero on the $j$-th qubit. Here we explicitly compute the expression in (\ref{locprep}) to obtain a quantity with a clear operational meaning. The final expression can be written as a sum of terms with ``bit-flips'':
\beq\label{localcost}
\begin{aligned}
C_{lhs}^{\text{state}} = 1 -  |\bra{0}V^{\dagger}(\theta)\ket{\psi_0}|^2 - &\left(\frac{n-1}{n}\right)\sum\limits_{j=1}^n| \bra{0}X_jV^{\dagger}(\theta)\ket{\psi_0} | ^2-
  \left(\frac{n-2}{n}\right)\sum\limits_{j<k}|\bra{0}X_j X_kV^{\dagger}(\theta)\ket{\psi_0} |^2\\
&- ... - \frac{1}{n}\sum\limits_{j<k<l<...}|\bra{0}X_j X_k X_l...V^{\dagger}(\theta)\ket{\psi_0}|^2 
\end{aligned} 
\eeq
%

%
The explicit expression for $C_{lhs}^{\text{state}}$ in (\ref{localcost}) allows us to use a local cost function on a \textit{classical} computer without running the circuit several times and taking shots. It was shown in \cite{cerezo2021cost} that if the circuit Ansatz $V(\theta)$ is sufficiently shallow then the variance of the gradient of the local cost function vanishes polynomially in the number of qubits, instead of exponentially. One can therefore expect that using the cost function in (\ref{localcost}) will lead to faster convergence when applied to classical compilation of quantum circuits. However, the large number of terms appearing in (\ref{localcost}) increases the difficulty of calculating its gradient, as compared with the expression in (\ref{cost_state_prep}) - note that this is not the case for quantum compilation schemes where one runs the circuit in Figure \ref{local} instead of calculating each term in (\ref{localcost}) explicitly. In what follows, we will introduce a scheme to truncate the expression in (\ref{localcost}) such that its gradient can be calculated more easily. In particular, we will take only the first $k$ terms in equation (\ref{localcost}). This opens the question as to whether the truncated expression has a barren plateau. Below, we will take a particular example such that the cost function can be studied analytically and show exactly how each "bit-flip" term in (\ref{localcost}) affects the variance of the gradient.\\

The specific example that we use to demonstrate the effect of the ``bit-flip" terms was also studied in \cite{cerezo2021cost}. We take
\beq\label{ansatz_ex}
V(\theta) = \otimes_{j=1}^{n}e^{-i\theta_j \frac{\sigma_x}{2}}
\eeq
and
\beq\label{target_ex}
\ket{\psi_0}=\ket{0}
\eeq
Equations (\ref{cost_state_prep}) and (\ref{locprep}) become respectively:
\beq\label{global_cost_val}
C_{hs}^{state} = 1 - \prod\limits_{j=1}^{n}\cos^2\frac{\theta_j}{2}
\eeq
\beq\label{local_cost_val}
C_{lhs}^{state} = 1 - \frac{1}{n}\sum\limits_{j=1}^{n}\cos^2\frac{\theta_j}{2}
\eeq
As discussed in \cite{cerezo2021cost}, the variance of the expression in (\ref{global_cost_val}) is given by:
\beq\label{var_grad}
\text{Var}\left[\frac{\partial C_{hs}^{state}}{\partial \theta_j} \right] = \frac{1}{8}\left(\frac{3}{8}\right)^{n-1}
\eeq
whereas for the expression in (\ref{local_cost_val}) we get:
\beq
\text{Var}\left[\frac{\partial C_{lhs}^{state}}{\partial \theta_j}\right] = \frac{1}{8n^2}
\eeq
Let's define a new truncated cost function as follows:
\beq\label{cost_func_alpha}
\begin{aligned}
C_L^{(k)}(\alpha_1,...,\alpha_k) = &1 - |\bra{0}V^{\dagger}(\theta)\ket{\psi_0}|^2 - \alpha_1\sum\limits_{j=1}^n|\bra{0}X_jV^{\dagger}(\theta)\ket{\psi_0}|^2 \\ &-\alpha_2\sum\limits_{j<k}|\bra{0}X_jX_kV^{\dagger}(\theta)\ket{\psi_0}|^2
-\alpha_k\sum\limits_{j_1<...<j_k}|\bra{0}X_{j_1}...X_{j_k}V^{\dagger}(\theta)\ket{\psi_0}|^2
\end{aligned}
\eeq
The goal is now to tune $\alpha_1,...,\alpha_k$ to increase the variance of this truncated expression. Note that if we take $k = n-1$ and $\alpha_1,...,\alpha_k = \frac{n-1}{n},...,\frac{1}{n}$ we recover the original expression in (\ref{localcost}). We can understand the intuitive meaning behind (\ref{cost_func_alpha}) in terms of "bit-flips"; first note that the second term in (\ref{cost_func_alpha}) is equal to exactly $1$ at the optimal values of $\theta$ and all subsequent terms are zero. We can therefore understand this expression in the following way: the second term encourages the parameters $\theta$ to tend towards those that correspond to the target state $\ket{\psi_0}$, while all other terms bias $\theta$ towards states with increasing numbers of "bit-flip" errors, i.e. $X_j\ket{\psi_0}$ and $X_kX_j\ket{\psi_0}$, which act so as to increase the size of the target in regions of barren parameter space.
%
%
Let's now take $k=1$, in which case equation (\ref{cost_func_alpha}) becomes
\beq
C_L^{(1)}(\alpha_1) \equiv 1 - \prod\limits_{j=1}^{n}\cos^2\frac{\theta_j}{2} - \alpha_1\sum\limits_{i_1}\sin^2\theta_{i_1}\prod\limits_{j \neq i_1}\cos^2\theta_j
\eeq
By setting $\alpha_1 = \frac{1}{n}$ we get:
\beq
C_L^{(1)}\left(\alpha = \frac{1}{n}\right) = 1 - \frac{1}{n}\sum\limits_i\prod\limits_{j\neq i} \cos^2 \theta_j
\eeq
Similarly, we can define:
\beq
C_L^{(2)}(\alpha_1, \alpha_2) \equiv 1 - \prod\limits_{j=1}^{n}\cos^2\frac{\theta_j}{2} - \alpha_1\sum\limits_{i_1}\sin^2\theta_{i_1}\prod\limits_{j \neq i_1}\cos^2\theta_j - \alpha_2\sum\limits_{i_1<i_2}\sin^2\theta_{i_1}\sin^2\theta_{i_2}\prod\limits_{j\neq i_1, i_2}\cos^2\theta_j
\eeq
Now by taking $\alpha_1$ and $\alpha_2$ to be the solution to the set of linear equations:
\beq\label{lin_eqns}
\begin{aligned}
n \alpha_1 -\alpha_2 {n \choose 2} &= 1 \\
-\alpha_1 + \alpha_2(n-1) &= 0
\end{aligned}
\eeq
we get
\beq
C_L^{(2)} = 1 - \frac{2}{n(n-1)}\sum\limits_{i_1, i_2}\prod\limits_{j\neq i_1, i_2} \cos^2 \theta_j
\eeq
Now, we can see that each truncated local expression reduces the power of the variance in (\ref{var_grad}) by 1. In particular, we get:
\beq\label{var_cl1}
\text{Var}\left[\frac{\partial C_{L}^{(1)}}{\partial \theta_j} \right] \propto \left(\frac{3}{8}\right)^{n-2}
\eeq
and
\beq\label{var_cl2}
\text{Var}\left[\frac{\partial C_{L}^{(2)}}{\partial \theta_j} \right] \propto \left(\frac{3}{8}\right)^{n-3}
\eeq
In the general case $C_L^{(k)}$, we have $k$ independent variables $\alpha_1,...,\alpha_k$ and we will obtain $k$ linear equations similar to (\ref{lin_eqns}) such that we get:
\beq\label{var_clk}
\text{Var}\left[\frac{\partial C_{L}^{(k)}}{\partial \theta_j} \right] \propto \left(\frac{3}{8}\right)^{n-k-1}
\eeq
We have thus shown that there exist coefficients $\alpha_1,...,\alpha_k$ such that the exponent that appears in the variance of the gradient of the truncated cost function in (\ref{cost_func_alpha}) is increased by $k$ with respect to the global cost function. This has practical implications as we will demonstrate in section \ref{sec:results}, where it will be shown that the convergence of (\ref{cost_func_alpha}) with $k=1$ is signicantly better than (\ref{hstest}). We can understand this behaviour from the analytical results in equations (\ref{var_cl1}) to (\ref{var_clk}). We conjecture that such results will hold for more general Ansatz and target states - see section \ref{sec:results}. In section \ref{sec:algorithm} we will present our method to update the $\alpha_1,...,\alpha_k$ over the course of the optimisation.\\

We note that we need to extend this analysis further in order to apply it to full approximate quantum compiling, i.e. where the target is a sub-quantum circuit that does not necessarily appear at the beginning of the algorithm, in which case the target to approximate is a matrix instead of a vector. In this case, a natural way to extend the cost function in (\ref{localcost}) is:
\beq\label{localcostmat}
\begin{aligned}
C_{lhs}^{\text{circuit}} = 1 - \frac{1}{d^2} [\  &|\Tr(V^{\dagger}U)|^2 + \left(\frac{n-1}{n}\right)\sum\limits_{j=1}^n |\Tr(X_j V^{\dagger}U)|^2 + \left(\frac{n-2}{n}\right)\sum\limits_{j<k}|\Tr(X_j X_k V^{\dagger}U)|^2\\
&+ ... + \frac{1}{n}\sum\limits_{j<k<l<...}|\Tr(X_j X_k X_l... V^{\dagger}U)|^2 ]
\end{aligned}
\eeq
In section \ref{sec:results}, we will present results showing how (\ref{localcostmat}) improves the convergence as compared to the Hilbert-Schmidt test in equation (\ref{hstest}).

\section{Algorithm}\label{sec:algorithm}

We now discuss the scheme we have developed to update the coefficients $\alpha$ appearing in equation (\ref{cost_func_alpha}) during the optimisation. Consider the cost function $C_L^{(1)}(\alpha)$:
\beq\label{cost_func_alpha2}
C_L^{(1)}(\alpha) = 1 - |\bra{0}V^{\dagger}(\theta)\ket{\psi_0}|^2 -
\alpha\sum\limits_{j=1}^n| \bra{0}X_j V^{\dagger}(\theta)\ket{\psi_0}|^2
\eeq
The parameter $\alpha$ is a function of the number of iterations, i.e. $\alpha \equiv \alpha(k)$. The first, most straightforward scheme that we implement is to gradually decrease $\alpha$ with increasing $k$ and hence reduce the influence of the final term in (\ref{cost_func_alpha2}) as the cost function approaches zero. In particular, we take $\alpha(k) = \sqrt{C_L^{(1)}(k)}$ and $\alpha(0) = 1$. Note, it can be shown that $C_L^{(1)}(\alpha)$ is always non-negative, even when $\alpha=1$, because $\ket{0}$ and all ``bit-flipped" states constitute an orthogonal basis. Intuitively, the weighting scheme acts such as to tighten a grip around the state $\ket{0}$ as $C_L^{(1)}(\alpha)$ approaches zero.  Eventually, when $C_L^{(1)}(\alpha) \approx 0$, all the flip terms are switched off and we arrive at the desired solution $V(\theta)\ket{0} \approx \ket{\psi_0}$. In \ref{weight_scheme} we discuss this weighting scheme in more detail, as well as our implementation of a fast-gradient approach based on \cite{madden2022sketching} which allows us to do experiments with up to 24 qubits. We also developed a number of techniques to improve performance, in particular we use a ``surrogate model" which speeds up the calculation of the gradient - we discuss these aspects of the algorithm in \ref{surrogate_mod}.

\section{Results}\label{sec:results}

\subsection{State Preparation}

\begin{figure}[!htb]\centering
\centering
\frame{\includegraphics[scale=0.4]{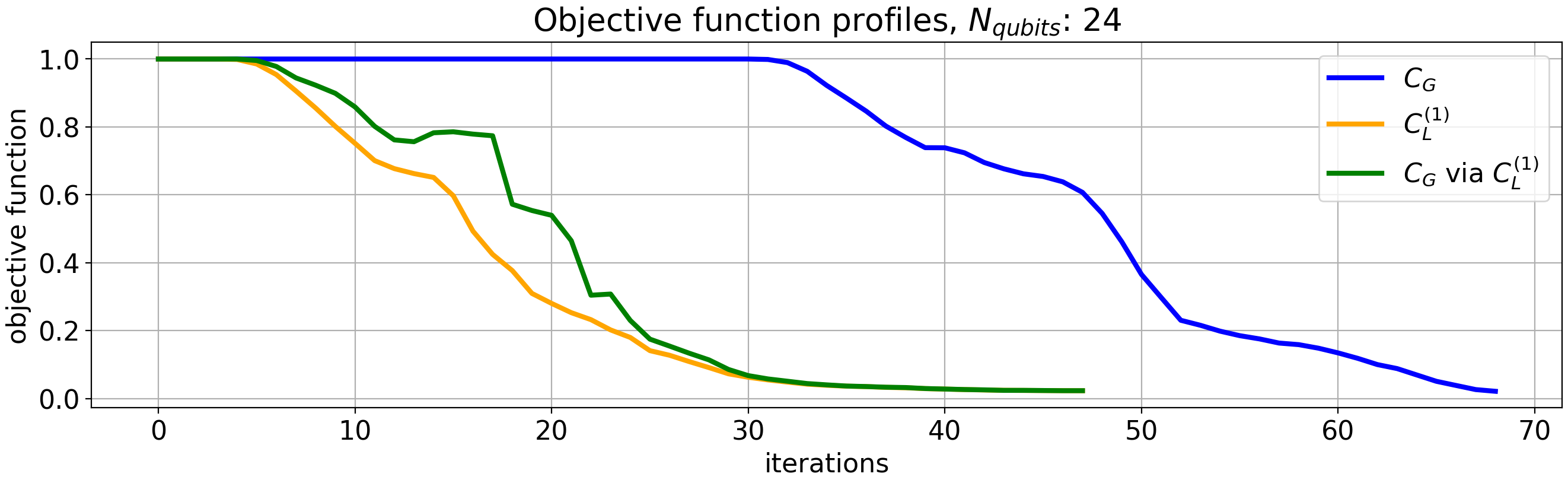}}
\caption{Influence of the flip terms on convergence rate. Not only does the local cost function $C_L^{(1)}$ converge at a much faster rate than $C_G$, but so too does $C_G$ when evaluated at the parameters $\theta$ that are found by optimising for $C_L^{(1)}$ - see the green curve ``$C_G$ via $C_L^{(1)}$''.}
\label{fig:lhs-vs-no-lhs}
\end{figure}

We now consider the results of our experiments on approximate state preparation using the localised cost functions discussed in previous sections. We take the circuit Ansatz to be that discussed in section \ref{sec:param_circ}. In the notation defined in Figure \ref{paramcircuit} and the surrounding text, we consider the case with $n=24$ qubits, $b=1$ repetitions and $l=1$ layers. Note that in this notation, Figure \ref{paramcircuit} corresponds to $n=4$, $b=3$ and $l=2$. In our case, there are therefore $24$ qubits and $24$ \cnbls connecting every couple of adjacent qubits. The target state $\ket{\psi_{0}}$ was generated from a circuit with the same $CNOT$ structure but parameterized by random angles. Figure \ref{fig:lhs-vs-no-lhs} clearly demonstrates that the objective function without the ``bit-flip" terms $C_G$ performs much worse than $C_L^{(1)}$ which takes into account the bit-flip terms.\\

It was observed that second-order, quasi-Newton \texttt{L-BFGS} method terminates prematurely if it does not detect any progress after a small number of iterations. In order to circumvent this problem, we employ two-stage optimization. Firstly, the first-order gradient-based \texttt{Adam} optimizer reduces the objective function from $1$ to a value below $0.9$, then \texttt{L-BFGS} does the rest of the job.

\subsection{Circuit preparation}

\begin{figure}[h]
\centering
\begin{tikzpicture}[scale = 1.5]
\draw[gray, thick] (0,1.13) -- (0,0.13);
\filldraw[black] (0,0.13) circle (2pt);
\draw[black] (0,1) circle (4pt);
\draw[gray, thick] (-0.14, 1) -- (0.14,1);

\draw[gray, thick] (-1.5,1) -- (1.1,1);
\draw[gray, thick] (-0.7,0.13) -- (1.1,0.13);

\node at (-1.1,0.13) {\footnotesize $R_z(-\frac{\pi}{2})$};

\draw[gray, thick] (-1.5,-0.17) -- (-1.5, 0.43);
\draw[gray, thick] (-0.7,-0.17) -- (-0.7, 0.43);
\draw[gray, thick] (-1.5, 0.43) -- (-0.7, 0.43);
\draw[gray, thick] (-1.5,-0.17) -- (-0.7,-0.17);

\draw[gray, thick] (1.1,0.7) -- (1.1,1.3);
\draw[gray, thick] (1.9,0.7) -- (1.9,1.3);
\draw[gray, thick] (1.1,1.3) -- (1.9,1.3);
\draw[gray, thick] (1.1,0.7) -- (1.9,0.7);

\draw[gray, thick] (1.1,-0.17) -- (1.1, 0.43);
\draw[gray, thick] (1.9,-0.17) -- (1.9, 0.43);
\draw[gray, thick] (1.1, 0.43) -- (1.9, 0.43);
\draw[gray, thick] (1.1,-0.17) -- (1.9,-0.17);

\draw[gray, thick] (1.9,1) -- (6.5,1);
\draw[gray, thick] (1.9,0.13) -- (4,0.13);

\node at (1.5,1) {$R_y(\theta_1)$};
\node at (1.5,0.13) {$R_y(\theta_3)$};

\draw[gray, thick] (3,1) -- (3,0);
\filldraw[black] (3,1) circle (2pt);
\draw[black] (3,0.13) circle (4pt);

\node at (4.4,0.13) {$R_y(\theta_3)$};

\draw[gray, thick] (4,-0.17) -- (4, 0.43);
\draw[gray, thick] (4.8,-0.17) -- (4.8, 0.43);
\draw[gray, thick] (4, 0.43) -- (4.8, 0.43);
\draw[gray, thick] (4,-0.17) -- (4.8,-0.17);

\draw[gray, thick] (4.8,0.13) -- (7.8,0.13);

\draw[gray, thick] (5.8,1.13) -- (5.8,0.13);
\filldraw[black] (5.8,0.13) circle (2pt);
\draw[black] (5.8,1) circle (4pt);
\draw[gray, thick] (5.66, 1) -- (5.94,1);

\draw[gray, thick] (6.5,0.7) -- (6.5,1.3);
\draw[gray, thick] (7.3,0.7) -- (7.3,1.3);
\draw[gray, thick] (6.5,1.3) -- (7.3,1.3);
\draw[gray, thick] (6.5,0.7) -- (7.3,0.7);

\draw[gray, thick] (7.3,1) -- (7.8,1);

\node at (6.9,1) {$R_z(\frac{\pi}{2})$};

\end{tikzpicture}
\caption{Implementation of two site operator $e^{i(\alpha \sigma^x \otimes \sigma^x + \beta \sigma^y \otimes \sigma^y + \gamma \sigma^z \otimes \sigma^z)}$ as a quantum circuit. We have the correspondences $\theta = \frac{\pi}{2}-2\gamma$, $\phi = 2\alpha-\frac{\pi}{2}$ and $\lambda=\frac{\pi}{2}-2\beta$. The Hamiltonian in (\ref{hxxz}) corresponds to the case $\alpha = \beta = \gamma = dt$}\label{trotter_circ}
\end{figure}
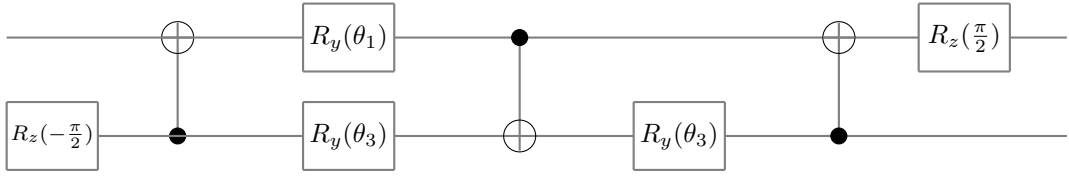

We now consider the results for full approximate quantum compiling for which the target is a matrix instead of a vector. As discussed in section \ref{sec:barren}, the truncated local cost function expression permits a natural extension from (\ref{localcost}) to (\ref{localcostmat}). In \cite{madden2022best}, full approximate compiling was applied to random unitaries. Here we consider the approximate compilation of circuits used to simulate time evolution of quantum systems. The motivation for considering these circuits is that they are approximations of the exact time evolution operator - to be discussed in detail below - and thus provide a useful benchmark against which we can compare the accuracy of our compilation. Time evolution of quantum systems is governed by the Schr{\"o}dinger equation:
\beq\label{schrod_eq}
 \ket{\psi(t)}= e^{-iHt} \ket{\psi(0)}
\eeq
where $\ket{\psi(0)}$ is the wavefunction at time $t=0$, and $H$ is the Hamiltonian of the system under study. We will consider the XXX Hamiltonian - a paradigmatic model for quantum magnetism - written explicitly in terms of Pauli matrices $\sigma^x, \sigma^y$ and $\sigma^z$ as:
\begin{equation}\label{hxxz}
    H_{XXZ} = -\sum_{\ell=0}^{L-1}\left( \sigma^x_\ell \sigma^x_{\ell+1} + \sigma^y_\ell \sigma^y_{\ell+1} + \sigma^z_\ell \sigma^z_{\ell+1}  \right),
\end{equation}
The time evolution operator $U(t)\equiv e^{-iHt}$ can be executed as a quantum circuit via Trotterisation which we now explain. We define $h_{\ell, \ell+1} = \left(\sigma^x_\ell \sigma^x_{\ell+1} + \sigma^y_\ell \sigma^y_{\ell+1} + \sigma^z_\ell \sigma^z_{\ell+1}\right)$ such that the Hamiltonian in (\ref{hxxz}) becomes $H_{XXZ} = H_1 + H_2$ where $H_1 = -\sum\limits_{\ell\text{ odd}} h_{\ell, \ell+1}$ and $H_2 = \sum\limits_{\ell\text{ even}} h_{\ell, \ell+1}$. Note that all operators in a given sum commute with all other operators in their respective sums. We then define the Trotterised time evolution operator $\mathcal{U}_{\text{trot}}(dt)$ in the following way:
\beq\label{utrott}
    \mathcal{U}_{\text{trot}}(dt) = \prod_{j=0}^{L/2-1}U_{2j, 2j+1}(dt)\prod_{j=1}^{L/2-1}U_{2j-1, 2j}(dt) = e^{-iH_{XXZ}dt} + O(dt^2)
\eeq
where $U_{j k}(dt) = e^{-ih_{j k}dt}$. The exact time evolution operator $U(t)$ is thus approximated by $m$ repeated applications of $\mathcal{U}_{\text{trot}}(dt=\frac{t}{m})$, i.e. $U(t) \approx \mathcal{U}^m_{\text{trot}}(dt=\frac{t}{m})$. As discussed in \cite{smith2019simulating}, each $U_{j k}(dt)$ appearing in (\ref{utrott}) can be implemented by the quantum circuit in Figure \ref{trotter_circ}. Equation (\ref{utrott}) is a first order expression, one can also consider second and higher order Trotter operators \cite{layden2022first} but with increased circuit depth. Note that $\mathcal{U}_{\text{trot}}(dt)$ is exact if one considers a time dependent Hamiltonian in the Floquet formalism, thus allowing for highly accurate quantum simulations \cite{keenan2022evidence}.\\

We now compare the fidelity of the operator we find via approximate quantum compiling with the exact time evolution operator vs the fidelity of the Trotter operator vs the time evolution operator. We note that a number of other techniques have been developed recently that aim to find short depth quantum circuits that implement time evolution \cite{gibbs2021long, cirstoiu2020variational, zoufal2021error, bharti2021quantum}. The fidelity of two matrix operators $U$ and $V$ of dimension $d \times d$ is defined as:
\beq
f(U,V) = \frac{1 + \frac{|\Tr V^{\dagger} U|^2}{d}}{d+1}
\eeq
\begin{figure}[!htb]\centering
	   \centering
     \frame{\includegraphics[scale=0.5]{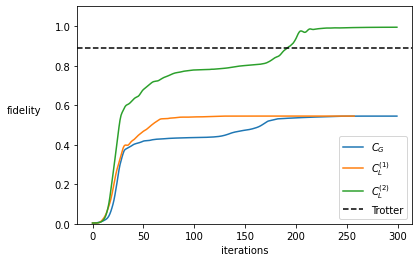}}
     \caption{Approximate compilation of time evolution operator on 8 qubits - the horizontal dotted line corresponds to the fidelity that one obtains with the standard first order Trotter decomposition with two Trotter steps and with $dt=0.2$.}\label{8q}
\end{figure}
In Figure \ref{8q} we plot the fidelity of the approximate quantum compiling operator with the exact time evolution operator vs the number of iterations in the optimisation procedure. We can see that both the global cost function $C_G$ defined in (\ref{hstest}) and the partially localised cost function defined in (\ref{cost_func_alpha}) with $k=1$ reach a barren plateau and converge to an operator that has poor fidelity with the target. However, we can see that $C_L^{(2)}$ performs much better and not only does it escape from the barren plateau, but achieves a fidelity that is higher than that of the Trotter circuit.

\section{Discussion}\label{sec:discussion}
In this work we have developed a set of techniques that significantly improve the convergence of classically assisted quantum compilation algorithms. Just like the local cost functions considered in \cite{khatri2019quantum, cerezo2021cost}, we have demonstrated in practice that our truncated local cost functions and their gradients (see e.g. equations (\ref{cost_func_alpha}) and (\ref{localcostmat})) can be feasibly calculated on a classical computer while also being powerful enough to escape barren plateaus that appear in both approximate state preparation and full circuit approximation - see Figures \ref{fig:lhs-vs-no-lhs} and \ref{8q}. As discussed in the introduction, there are a number of circumstances where this classically assisted approach can be useful. One might for example be interested in compiling sub-circuits that appear in a given quantum algorithm, or alternatively one might be interested in finding the most efficient way to generate a weakly entangled state on a quantum computer \cite{rudolph2022decomposition}; one can then use this state generated by a shallow circuit as the input for a quantum algorithm \cite{rudolph2022synergy}. In the latter case, one can use Tensor Networks to represent the state classically and thus one can apply the techniques developed here on large numbers of qubits - see our upcoming work.

\newpage

\bibliography{references}{}
\bibliographystyle{unsrt}

\newpage

\appendix
\section{Weighting scheme}\label{weight_scheme}

As discussed in section \ref{sec:algorithm}, the parameter $\alpha$ in equation (\ref{cost_func_alpha2}) is gradually decreased over the optimisation, enforcing $V^{\dagger}(\theta)\ket{\psi_0}$ to have a large projection on $\ket{0}$ while progressing mostly inside the sub-space spanned by states $\ket{0}$ and $\{X_j \ket{0} | 1 \le j \le n\}$. One such approach - as mentioned in the main text - is to take $\alpha(k) = \sqrt{C_L^{(1)}(k)}$ and $\alpha(0) = 1$. We find that slightly modifying the expression for $\alpha(k)$ to e.g.:
\begin{equation}\label{eq:weight}
w^{k} = 0.1 \cdot \sqrt{C_{lhs}^{k-1}} + 0.9 \cdot w^{k-1}, \quad w^0 = 1.
\end{equation}
can enhance the optimisation procedure due to the reduced noise in the gradient profile. To increase speed, we apply a fast gradient approach based on \cite{madden2022sketching} by using a more economical matrix-vector multiplication. However this approach must be modified in order to apply it to our case; the main bottleneck is that the gradient of every term in the sum in equation (\ref{cost_func_alpha2}) should be computed separately and then all of them summed up. This causes issues with speed when considering sizes of up to 24 qubits. To handle this, we designed and implemented a surrogate model approach which we describe in \ref{surrogate_mod}.

\section{Surrogate Model}\label{surrogate_mod}

In the second formulation, instead of summing up squared projections onto individual flip-states, we minimize the maximal projection of $V^{\dagger}(\theta)\ket{\psi_0}$ onto the sub-space spanned by states $\ket{0}$ and $\{X_j \ket{0} | 1 \le j \le n\}$, which we call a \textit{surrogate model}:
\begin{equation}\label{eq:fobj-sur}
\begin{aligned}
C_{lhs}^{surr}(\theta; \alpha) &= \min_{{\boldsymbol\beta}, \|{\boldsymbol\beta}\|^2=1} \Bigg\{
1 - (1 - \alpha) |\bra{0}V^{\dagger}(\theta)\ket{\psi_0}|^2 - \\
& \qquad\qquad\qquad - \alpha\left|\left(\beta_0^* \bra{0} + \sum\limits_{j=1}^n \beta_j^* \bra{0}X_j\right)
V^{\dagger}(\theta)\ket{\psi_0}\right|^2
\Bigg\},
\end{aligned}
\end{equation}
where on every iteration the objective is minimized over the parameters ${\boldsymbol\beta}$ by taking the maximal projection onto the sub-space of states $\ket{0}$ and $\{X_j \ket{0} | 1 \le j \le n\}$ (which is the same as \textit{maximizing} the third term) before optimization over the angular parameters $\theta$. It is straightforward to show that the following normalized $\beta$'s, $\|{\boldsymbol\beta}\|^2 = 1$, maximize the projection: $\beta_0 \sim \bra{0}V^{\dagger}(\theta)\ket{\psi_0}$ and $\beta_j \sim \bra{0}X_jV^{\dagger}(\theta)\ket{\psi_0}$\footnote{Really, let us define $y_0$ = $\bra{0}V^{\dagger}(\theta)\ket{\psi_0}$ and $y_j$ = $\bra{0}X_jV^{\dagger}(\theta)\ket{\psi_0}$. Then the last term in (\ref{eq:fobj-sur}) takes the form of squared module of a dot product: $|\sum\nolimits_{i=0}^n \beta_i^* y_i|^2$, which is maximized when $\beta_i \sim y_i$.}.

On every iteration, we compute the latter dot products for $\beta$'s and normalize the vector ${\boldsymbol\beta}$, then substitute $\beta$-parameters in the last term in (\ref{eq:fobj-sur}) and compute the \textit{composite} state $\beta_0^* \bra{0} + \sum\nolimits_{j=1}^n \beta_j^* \bra{0}X_j$. Afterwards, we proceed with optimization over the parameters $\theta$ as usually. The computational gain is obvious --- the gradient is computed for just two dot products instead of $n+1$ ones, and the gradient calculation is known to be the most time-consuming operation. 

For the additional speed up we compute the state $V^{\dagger}(\theta)\ket{\psi_0}$ at the beginning of each optimization iteration given the current parameters $\theta$ and use it for computation of both, the objective function and its gradient, see \cite{madden2022sketching} for details. This substantially reduces the processing time.

Despite the advantage of surrogate model, adding up individual states to get a composite one, $\beta_0^* \bra{0} + \sum\nolimits_{j=1}^n \beta_j^* \bra{0}X_j$, can be challenging if we want to generalize the framework from using an ordinary vectors to matrix-product state (MPS) format for scalability. In view of future generalization, we devised even more simple surrogate model, where instead of composite state we pick up one of flipped states or $\ket{0}$, whichever gives the maximal projection onto $V^{\dagger}(\theta)\ket{\psi_0}$, and solely use it in the last term: 
\begin{equation}\label{eq:fobj-sur-max}
C_{lhs}^{max}(\theta; \alpha) = 
1 - (1 - \alpha) |\bra{0}V^{\dagger}(\theta)\ket{\psi_0}|^2 - 
\alpha\left|\bra{j_{max}}V^{\dagger}(\theta)\ket{\psi_0}\right|^2,
\end{equation}
where $j_{max} = 0$, if $|\bra{0}V^{\dagger}\ket{\psi_0}|$ gives the largest projection, or $j_{max} = j^{\prime}$, $1 \le j^{\prime} \le n$, if $|\bra{0}X_{j^{\prime}}V^{\dagger}\ket{\psi_0}|$ has the greatest magnitude.

One more feature is required to make the objective (\ref{eq:fobj-sur-max}) viable. If component of the largest projection is altered too often, this could contribute a strong disturbance into the optimization process. In order to alleviate that adverse effect, we employ the idea of hysteresis. Namely, the former leading state $\ket{j_{max}}$ only gives up if its projection onto $V^{\dagger}(\theta)\ket{\psi_0}$ is $10\%$ lower than the projection of the currently leading state, otherwise optimization proceeds with the former leader.

\end{document}